# Predicting Risk of Pulmonary Fibrosis Formation in PASC Patients

Wanying Dou*, Gorkem Durak*, Koushik Biswas*, Ziliang Hong*, Andrea Mia Bejar*, Elif Keles*, Kaan Akin†,
Sukru Mehmet Erturk†, Alpay Medetalibeyoglu†, Marc Sala*, Alexander Misharin*, Hatice Savas*,
Mary Salvatore‡, Sachin Jambawalikar‡, Drew Torigian§, Jayaram K. Udupa‡, Ulas Bagci*
*Northwestern University, Chicago, IL, USA
†Istanbul University, Istanbul, Turkey
‡Columbia University, New York City, NY, USA
§University of Pennsylvania, Philadelphia, PA, USA
Correspondence: ulas.bagci@northwestern.edu

*Abstract*—While the acute phase of the COVID-19 pandemic has subsided, its long-term effects persist through Post-Acute Sequelae of COVID-19 (PASC), commonly known as Long COVID. There remains substantial uncertainty regarding both its duration and optimal management strategies. PASC manifests as a diverse array of persistent or newly emerging symptoms—ranging from fatigue, dyspnea, and neurologic impairments (e.g., brain fog), to cardiovascular, pulmonary, and musculoskeletal abnormalities—that extend beyond the acute infection phase. This heterogeneous presentation poses substantial challenges for clinical assessment, diagnosis, and treatment planning. In this paper, we focus on imaging findings that may suggest fibrotic damage in the lungs, a critical manifestation characterized by scarring of lung tissue, which can potentially affect long-term respiratory function in patients with PASC. This study introduces a novel multi-center chest CT analysis framework that combines deep learning and radiomics for fibrosis prediction. Our approach leverages convolutional neural networks (CNNs) and interpretable feature extraction, achieving 82.2% accuracy and 85.5% AUC in classification tasks. We demonstrate the effectiveness of Grad-CAM visualization and radiomics-based feature analysis in providing clinically relevant insights for PASC-related lung fibrosis prediction. Our findings highlight the potential of deep learning-driven computational methods in enabling early detection and risk assessment of PASC-related lung fibrosis—presented for the first time in the literature.

## I. INTRODUCTION

### A. Background and Motivation

Post-acute sequelae of COVID-19 (PASC), commonly known as long COVID, affects over 65 million people globally [1], with economic consequences exceeding $50 billion annually in lost labor productivity in the United States

This study is supported by NIH R01 HL171376.

alone [2]–[4]. The condition presents with a wide spectrum of symptoms involving multiple organ systems, with respiratory complications posing particular concern for long-term health outcomes. Common symptoms including fatigue, cognitive impairment, respiratory difficulties, and psychological disturbances can persist for months or even years, significantly impacting daily functioning and overall well-being [5]–[9]. Among these complications, pulmonary fibrosis—characterized by progressive and potentially irreversible scarring of lung tissue—represents one of the most significant long-term consequences, underscoring the importance of early detection and timely intervention.

### B. Current Challenges and Related Work

Early research on PASC has identified diverse symptoms, potential biomarkers, and links to immune dysfunction. Clinical observations suggest delayed neurological symptoms and persistent immune activation; however, definitive diagnostic tools are still in development [1], [10]. Current clinical guidelines advocate for a holistic approach that integrates clinical evaluation with non-invasive diagnostic tools—such as chest computed tomography (CT), cardiac magnetic resonance imaging (CMR), pulmonary function tests, and biomarker analysis—to support the identification and monitoring of PASC-related complications [11]. Computational approaches including deep learning (DL), machine learning (ML), and radiomics analysis have emerged as powerful tools for addressing these diagnostic challenges, offering unprecedented capabilities for pattern recognition in complex medical datasets. These methods can analyze clinical, imaging, and laboratory data to uncover subtle patterns often overlooked by traditional diagnostics. As PASC cases continue to rise and symptom presentations remain heterogeneous, AI-driven computational models offer a promising path for early identification, risk stratification, and prediction of disease trajectories [12]. While advanced computing techniques have been widely applied to electronic health records (EHR) to identify patterns and risk factors associated with PASC [10], [13]–[15], their use in imaging-based PASC prediction remains limited-primarily due to the lack of large, diverse imaging datasets.

Despite data constraints, imaging studies have yielded crucial insights into the pathophysiology of PASC. For instance, CT imaging has detected persistent abnormalities, such as fibrosis-like changes, in some patients months after infection. Even among recovered individuals, chest abnormalities can persist, suggesting potential long-term pulmonary involvement in PASC [16], [17]. However, the intersection of advanced imaging analytics—particularly deep learning and radiomics—with PASC-specific pulmonary manifestations represents a critical yet underexplored research frontier.

*C. Contributions*

Our research presents an integrated computational approach combining deep learning and radiomics-based analysis on a multi-center CT imaging dataset to enhance assessment and prediction of PASC-related lung abnormalities. Our work makes the following key contributions:

- Introduction of a novel multi-center CT imaging dataset specifically curated for PASC-related fibrosis analysis[1].
- Development of a hybrid deep learning and radiomics-based framework achieving state-of-the-art classification performance.
- Comprehensive evaluation of both deep learning and radiomics-based approaches for fibrosis prediction with associated interpretable visualizations.

To the best of our knowledge, this work represents the first attempt to address the prediction of imaging findings indicative of pulmonary fibrosis in PASC patients using computational methods. The ability to identify patients at risk of developing permanent lung damage has significant clinical implications, potentially enabling earlier interventions and personalized treatment strategies for this emerging post-pandemic health challenge.

## II. METHODS

Given the diverse manifestations of PASC in lung imaging, accurate classification remains a challenge and requires strong analytical approaches. To address this, our study employs two complementary strategies, which represent a novel approach in this domain: **(1)** Deep learning models to improve classification precision, supporting early diagnosis and targeted clinical interventions, and **(2)** Radiomics-based analysis to extract quantitative imaging characteristics for PASC-related lung abnormalities. **Figure 1** illustrates the overall workflow of our methodology, where multi-center CT images are analyzed through both deep learning and radiomic feature extraction pipelines. The deep learning approach utilizes convolutional neural network (CNN)-based classification models with

---

[1]available at request, and will be freely available after review process.

Class Activation Mapping (CAM) techniques—specifically gradient-weighted CAM (Grad-CAM), for model interpretability. Meanwhile, the radiomics approach quantifies the texture, intensity, shape, and wavelet-based features from the same images, offering complementary analytical insights.

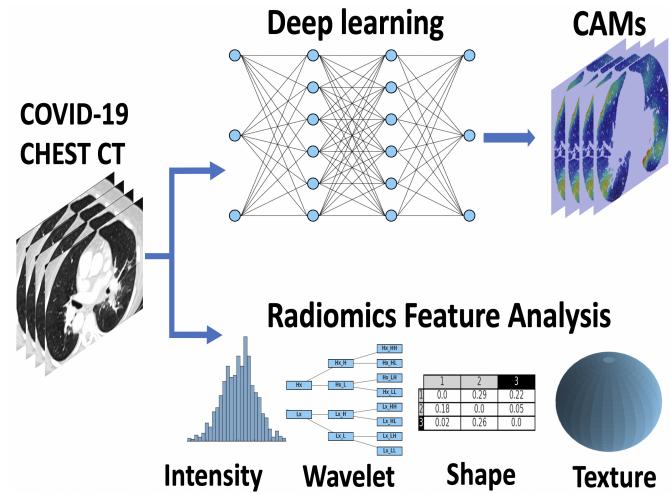

Fig. 1. Experimental workflow integrating deep learning for classification, Grad-CAM for visualization, and radiomics for feature extraction from multi-center CT images.

*A. Dataset and pre-processing*

Our dataset includes 347 chest CT scans: 233 from the COVID center of the University of Pennsylvania and 114 from multiple hospitals affiliated with the Istanbul School of Medicine in Turkey. Of these, 156 cases (44.9%) were positive for findings suggestive of pulmonary fibrosis (112 from Pennsylvania and 44 from Turkey), while 191 cases (55.1%) were negative (122 from Pennsylvania and 70 from Turkey). This balanced distribution across both centers ensures robust model training and evaluation. All cases were confirmed as COVID-19 positive through PCR testing, chest CT findings, and clinical evaluation. This dataset represents a diverse patient population, encompassing adults of varying ages and ethnic backgrounds, while also capturing imaging characteristics relevant to PASC-related findings. **Figure 2** presents representative images from the dataset.

A thoracic radiologist with eight years of experience evaluated all scans and categorized them according to persistent imaging abnormalities that may indicate permanent lung damage observed on follow-up chest CT scans taken at least two months after diagnosis. These expert evaluations were used as the reference truth labels for training and validating our models. The persistent imaging findings evaluated included ground-glass opacities, consolidation, reticulation, subpleural linear-curvilinear bands, parenchymal bands or micro-atelectasis, distortion of parenchymal architecture, honeycombing, traction bronchiectasis, and bronchial thickening or distortion.

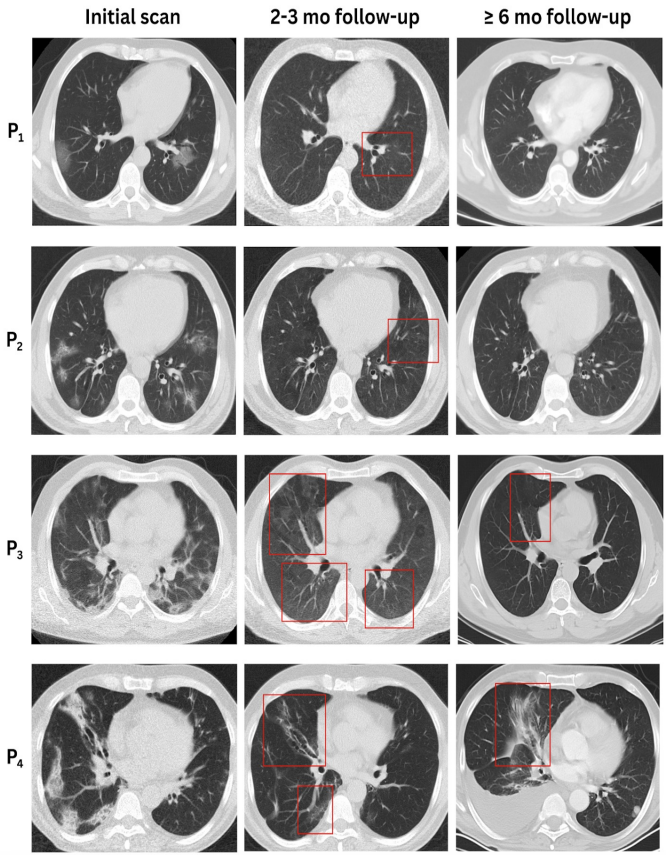

Fig. 2. Representative CT images of PASC cases at the initial scan, 2–3 month follow-up, and $\geq 6$ month follow-up. Red boxes highlight persistent imaging findings following COVID-19 pneumonia.

## B. Deep learning architecture and implementation

Our approach leverages multiple state-of-the-art CNN architectures to capture the complex imaging patterns associated with pulmonary fibrosis in PASC patients. While the CT data are inherently three-dimensional, we adopted a two-dimensional 2D approach for our deep learning models to optimize computational efficiency and maximize available training samples. Specifically, we extracted representative axial slices from each CT volume, focusing on sections containing lung tissue with potential fibrotic changes as identified by our radiologist. This slice-based approach allowed us to leverage well-established 2D CNN architectures while effectively capturing the radiological features indicative of pulmonary fibrosis. For each patient, multiple slices were extracted to ensure comprehensive representation of the lung volume, with the slice-level predictions subsequently aggregated using majority voting to determine the final patient-level classification. For rigorous evaluation, we employed a 10% test split as a held-out validation set and implemented five-fold cross-validation on the remaining 90% of the data. We systematically evaluated five carefully selected CNN architectures: DenseNet-121 [18], ResNet-18 [19], ResNet-34 [19], ResNeXt [20], and MobileNet V2 [21]. Each architecture was chosen for its unique characteristics and proven performance in medical imaging tasks. The training process was carefully optimized to ensure robust model performance. We employed a batch size of 2 and initialized the learning rate at 0.01, training each model for 100 epochs. To enhance generalization and prevent overfitting, we incorporated MixUp data augmentation techniques. The training was conducted on an Nvidia A100 GPU with automatic mixed precision (AMP) enabled, significantly optimizing memory efficiency and computational speed. We utilized Stochastic Gradient Descent (SGD) optimization with a momentum of 0.9 and a weight decay of 5e-4. The learning rate was dynamically adjusted using a cosine annealing scheduler to achieve optimal convergence [22].

## C. Radiomics feature extraction and analysis

The radiomics component of our framework involves the systematic extraction and analysis of quantitative imaging features using the PyRadiomics framework [23]. We extracted a comprehensive set of 107 radiomic features, categorized into several distinct groups to capture different aspects of the imaging phenotype. First-order statistical features (19 features) were computed to characterize the distribution of voxel intensities within the region of interest, providing insights into overall tissue density patterns. Shape-based features (26 features) were extracted to quantify both three-dimensional (16 features) and two-dimensional (10 features) geometric properties of the regions. The texture-based features (61 features) provided detailed information about spatial relationships and intensity patterns, comprising features from multiple matrices: Gray Level Co-occurrence Matrix (24 features), Gray Level Run Length Matrix (16 features), Gray Level Size Zone Matrix (16 features), and both Neighboring Gray Tone Difference Matrix and Gray Level Dependence Matrix (5 features each). To ensure consistency and minimize potential bias, all features were standardized using z-score normalization prior to model training. The feature extraction process was validated through extensive quality control measures to ensure reproducibility and reliability of the extracted features.

## III. EXPERIMENTS AND RESULTS

### A. Deep Learning Model Performance

The performance evaluation of our deep learning models revealed notable variations across different architectures (See **Table I**). DenseNet-121 emerged as the top performer in terms of accuracy, achieving 82.21% with an AUC of 83.84%. This superior performance can be attributed to its dense connectivity pattern, which promotes enhanced feature reuse and improved gradient flow during training. ResNet-18 demonstrated strong discriminative capability with the highest AUC of 85.46%, despite a slightly lower accuracy of 80.22%. ResNet-34 exhibited comparable performance with 81.66% accuracy and 84.70% AUC, suggesting that the additional depth compared to ResNet-18 did not necessarily translate to improved performance for this specific task. The ResNext architecture achieved moderate performance with

78.16% accuracy and 81.30% AUC, although MobileNet-V2 is notable for its computational efficiency, recorded the lowest performance with 63.16% accuracy and a 69.71% AUC. These findings indicate that the subtle and complex patterns of pulmonary fibrosis in PASC patients benefit from more sophisticated network architectures that can effectively capture hierarchical imaging features.

### B. Radiomics-Based Model Evaluation

In our radiomics analysis, we evaluated four different machine learning approaches for feature-based classification. **Table II** summarizes the results. The LASSO model demonstrated the strongest performance with an AUC of 72.54% and accuracy of 67.42%, highlighting its effectiveness in selecting relevant radiomic features while maintaining model interpretability. XGBoost showed comparable accuracy at 67.41% but with a lower AUC of 67.30%, suggesting slightly less robust discrimination ability. The Support Vector Machine (SVM) and Random Forest models achieved similar performance levels, with accuracies of 66.13% and 66.46% respectively, and AUCs of 65.72% and 66.17%. Although these performance metrics are lower than those achieved by the deep learning models, the radiomics approach offers crucial complementary value through its ability to identify specific quantifiable imaging biomarkers that may not be explicitly captured by neural networks, potentially enhancing clinical interpretability and providing biologically relevant insights.

### C. Model Interpretability and Visualization

To enhance the clinical utility of our models, we employed Gradient-weighted Class Activation Mapping (Grad-CAM) visualization techniques [24]. These visualizations provide crucial insights into the regions of the CT images that most significantly influence the model's predictions. In **Figure 3**, the Grad-CAM analysis revealed that our models consistently focused on areas of the lung showing characteristic patterns of fibrosis, including regions of ground-glass opacity, reticulation, and architectural distortion. The visualization results demonstrated remarkable concordance with radiological expertise, highlighting anatomical regions that radiologists typically prioritize during fibrosis assessment. This alignment supports the clinical validity of our models and provides a basis for their potential integration into clinical workflows.

TABLE I
PERFORMANCE OF DEEP LEARNING MODELS ON THE PASC DATASET (BEST RESULTS ARE BOLDED)

| Model | Accuracy (%) | AUC (%) |
|---|---|---|
| DenseNet-121 | **82.21** | 83.84 |
| ResNet-18 | 80.22 | **85.46** |
| ResNet-34 | 81.66 | 84.70 |
| ResNext | 78.16 | 81.30 |
| MobileNet-V2 | 63.16 | 69.71 |

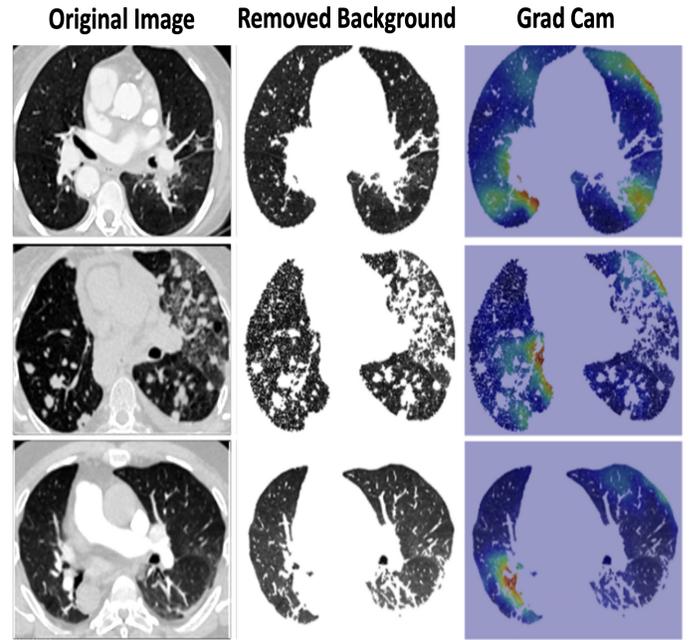

Fig. 3. Grad-CAM visualizations of the DenseNet-121 model applied to the newly collected PASC dataset.

TABLE II
PERFORMANCE OF MODELS USING RADIOMICS ON THE PASC DATASET (BEST RESULTS ARE BOLDED)

| Model | Accuracy (%) | AUC (%) |
|---|---|---|
| SVM | 66.13 | 65.72 |
| LASSO | **67.42** | **72.54** |
| XGBoost | 67.41 | 67.30 |
| Random Forest | 66.46 | 66.17 |

## IV. DISCUSSION

### A. Technical insights and model performance

Our comprehensive evaluation of both deep learning and radiomics-based approaches provided several important insights into the automated prediction of PASC patients likely to develop persistent lung damage. The strong performance of DenseNet-121 and ResNet architectures compared to lighter models suggests that the complexity of lung tissue damage patterns demands advanced feature extraction capabilities.

This finding is substantiated by our experimental results: DenseNet-121 achieved the highest accuracy (82.21%), while ResNet-18 showed the best discriminative power with an AUC of 85.46%. Analysis of our Grad-CAM visualizations (Figure 3) indicates that DenseNet-121's feature maps consistently emphasized subtle ground-glass opacities and early reticular patterns characteristic of developing fibrosis. This is likely due to its densely connected architecture, which enhances feature reuse across layers. Similarly, ResNet models were highly responsive to fine-grained textural changes in the lung parenchyma, preserving critical spatial information through skip connections that alleviate the vanishing gradient problem in deeper networks. These findings align with prior research, which has reported similar advantages of these architectures in identifying subtle pathological changes in medical imaging.

The radiomics-based models also contribute meaningful diagnostic value. Although their overall accuracy was lower than the deep learning models, their strength lies in the quantification of specific imaging features, offering insights that are particularly useful for clinical interpretation. The LASSO model's notable performance in radiomics analysis suggests that a relatively small subset of imaging features can effectively capture patterns associated with persistent damage.

*B. Limitations and challenges*

Despite the promising results of our computational approach, several important limitations must be acknowledged. First, while our multi-center dataset provides diversity in patient populations and imaging characteristics, further validation across additional healthcare systems and geographic regions is necessary to ensure broad generalizability. Second, the current approach focuses primarily on imaging features, while PASC manifestations often involve complex interactions between multiple physiological systems. Third, while persistent imaging findings are suggestive, longer-term follow-up imaging is needed to classify them as permanent. Fourth, we recognize methodological constraints regarding advanced deep learning architectures. A notable limitation in our study relates to the applicability of recent advanced architectures, such as Vision Transformers, foundation models, or state-of-the-art sequential models like Mamba. While these architectures have shown remarkable performance in general computer vision tasks, their application to our specific medical imaging context presents significant challenges, such as overfitting. The primary constraint is the limited dataset size compared to the massive datasets required for training such architectures effectively. Fifth, our experimental design deliberately chose training from scratch rather than utilizing transfer learning approaches with pre-trained weights. This decision was made to establish baseline performance specific to PASC-related fibrosis patterns without potential domain shift issues between natural images and our specialized CT imaging task. In future work, we plan to explore transfer learning strategies, including models pre-trained on large-scale chest CT datasets, which may further improve performance while reducing training data requirements.

*C. Conclusion*

This study presents a comprehensive framework for predicting the risk of pulmonary fibrosis in patients with PASC by analyzing chest CT imaging, marking the first such effort in the literature. The results suggest potential utility in supporting early detection and risk stratification of pulmonary complications related to PASC, utilizing machine learning and CT imaging.